\documentclass[conference]{IEEEtran}

\IEEEoverridecommandlockouts
\usepackage{cite}
\usepackage{amsmath,amssymb,amsfonts}
\usepackage{amsthm}
\usepackage{algorithmic}
\usepackage{graphicx}
\usepackage{textcomp}
\usepackage{xcolor}
\usepackage{hyperref}
\usepackage{makecell}
\usepackage{algorithm}
\usepackage{enumitem}
\usepackage{bbold}
\usepackage{algorithmic}
\usepackage{bbm}
\usepackage{quantikz} 

\usepackage{xcolor}
\usepackage[absolute,overlay]{textpos}
\newtheorem*{theorem*}{Theorem}

\hypersetup{
    colorlinks=true, 
    linkcolor=black,  
    citecolor=black, 
    filecolor=black, 
    urlcolor=black    
}

\def\BibTeX{{\rm B\kern-.05em{\sc i\kern-.025em b}\kern-.08em
    T\kern-.1667em\lower.7ex\hbox{E}\kern-.125emX}}
\begin{document}

\title{Special-Unitary Parameterization for\\ Trainable Variational Quantum Circuits \thanks{The views expressed in this article are those of the authors and do not represent the views of Wells Fargo. This article is for informational purposes only. Nothing contained in this article should be construed as investment advice. Wells Fargo makes no express or implied warranties and expressly disclaims all legal, tax, and accounting implications related to this article.}
}

\author{
\IEEEauthorblockN{
    Kuan-Cheng Chen\IEEEauthorrefmark{2}\IEEEauthorrefmark{3}\IEEEauthorrefmark{1},
    Huan-Hsin Tseng\IEEEauthorrefmark{4},
    Samuel Yen-Chi Chen \IEEEauthorrefmark{5},
    Chen-Yu Liu\IEEEauthorrefmark{6},
    Kin K. Leung\IEEEauthorrefmark{2}}

\IEEEauthorblockA{\IEEEauthorrefmark{2}Department of Electrical and Electronic Engineering, Imperial College London, London, UK}
\IEEEauthorblockA{\IEEEauthorrefmark{3}Centre for Quantum Engineering, Science and Technology (QuEST), Imperial College London, London, UK}
\IEEEauthorblockA{\IEEEauthorrefmark{4}Brookhaven National Laboratory, New York, USA}
\IEEEauthorblockA{\IEEEauthorrefmark{5}Wells Fargo, New York, USA}
\IEEEauthorblockA{\IEEEauthorrefmark{6}National Taiwan University, Taipei, Taiwan}
\IEEEauthorblockA{\IEEEauthorrefmark{1} Email: kuan-cheng.chen17@imperial.ac.uk}
}


\maketitle

\begin{abstract}
We propose SUN-VQC, a variational-circuit architecture whose elementary layers are single exponentials of a symmetry-restricted Lie subgroup, $\mathrm{SU}(2^{k}) \subset \mathrm{SU}(2^{n})$ with $k \ll n$. Confining the evolution to this compact subspace reduces the dynamical Lie-algebra dimension from $\mathcal{O}(4^{n})$ to $\mathcal{O}(4^{k})$, ensuring only polynomial suppression of gradient variance and circumventing barren plateaus that plague hardware-efficient ansätze. Exact, hardware-compatible gradients are obtained using a generalized parameter-shift rule, avoiding ancillary qubits and finite-difference bias. Numerical experiments on quantum auto-encoding and classification show that SUN-VQCs sustain order-of-magnitude larger gradient signals, converge 2--3$\times$ faster, and reach higher final fidelities than depth-matched Pauli-rotation or hardware-efficient circuits. These results demonstrate that Lie-subalgebra engineering provides a principled, scalable route to barren-plateau-resilient VQAs compatible with near-term quantum processors.
\end{abstract}


\section{Introduction}
\label{sec:intro}

Recent advances in quantum hardware, particularly the emergence of noisy intermediate-scale quantum (NISQ) devices, have catalyzed new strategies to harness quantum resources for solving classically intractable problems\cite{preskill2018quantum,bharti2022noisy}. Among these strategies, variational quantum algorithms (VQAs) have emerged as a leading framework, combining parameterized quantum circuits with classical optimizers to minimize task-specific cost functions\cite{cerezo2021variational,bittel2021training,lubasch2020variational,bonet2023performance}. These hybrid routines are not only well-suited to near-term hardware limitations due to their shallow circuit depth\cite{shang2023schrodinger}, but also benefit from the flexibility to incorporate physical symmetries and structural priors into the ansätze design\cite{meyer2023exploiting}. More fundamentally, VQAs embody a shift toward a quantum-native computational paradigm, where learning\cite{chen2020variational,chen2022variational,skolik2022quantum,du2022distributed,liu2024learning,chen2024compressedmediq}, optimization\cite{moll2018quantum,barkoutsos2020improving,bonet2023performance,chen2024noise,chen2025learning}, and information encoding\cite{khoshaman2018quantum,schuld2021effect} are realized through the continuous evolution of unitary operators within the geometry of Hilbert space\cite{yu2024shedding}. This formalism offers a unifying foundation for developing general-purpose quantum learning systems that are intrinsically tied to the algebraic structure of quantum mechanics.

This conceptual framework has found particular relevance in the field of quantum machine learning (QML), where variational circuits serve as the backbone of quantum models for classification\cite{blance2021quantum,chen2024quantum,chen2024quantum2,chen2024consensus,ma2025robust}, regression\cite{lin2024quantum,hsu2025quantum,liu2025programming,chen2025toward}, generative modeling\cite{chen2025differentiable,liu2025quantum,chen2025distributed,liu2024federated}, and reinforcement learning\cite{chen2020variational,liu2024qtrl,chen2024differentiable, chen2022variationalQRL,chen2025quantum}. In such models, quantum circuits act as parameterized learning maps that leverage entanglement, interference, and high-dimensional Hilbert space structure to potentially surpass the representational capacity of classical networks\cite{benedetti2019parameterized}. Recent developments have pushed this further into the domain of quantum artificial intelligence , wherein quantum models exhibit adaptive behavior, generalization, and structural inference across diverse data modalities. From quantum kernel estimation and variational autoencoders to reinforcement learning agents embedded in quantum environments, the variational quantum circuit (VQC) has become the central object of study. Understanding the geometric and algebraic properties of these circuits is therefore essential to enabling scalable, expressive, and trainable quantum learning systems.

However, the promise of VQAs and their applications in QML is severely challenged by a fundamental issue of trainability known as the \textit{barren plateau} (BP) phenomenon\cite{larocca2025barren}. As the number of qubits increases, the gradients of the cost function with respect to variational parameters can vanish exponentially, making gradient-based optimization statistically intractable. This behavior has been theoretically linked to the algebraic properties of the ansätze, specifically the dimension of the dynamical Lie algebra (DLA) $ \mathfrak{g} $ generated under the circuit’s adjoint action, with gradient variance scaling inversely with $ \dim \mathfrak{g} $\cite{meyer2023exploiting}. In typical hardware-efficient ansätze, composed of single-qubit Pauli rotations and entangling gates, $ \mathfrak{g} $ becomes isomorphic to $ \mathfrak{su}(2^n) $, leading to exponential growth in the accessible state manifold. Paradoxically, the same expressivity that allows a quantum circuit to explore high-dimensional solution spaces becomes detrimental to trainability when not carefully controlled\cite{cerezo2021cost}.

To overcome this limitation, we propose a theoretical framework for symmetry-informed variational quantum learning, which we call \textbf{SUN-VQC}. This approach restricts the circuit’s evolution within a smaller Lie subgroup $\mathrm{SU}(2^k)$ of the special unitary group $\mathrm{SU}(2^n)$ for some $k < n$, chosen to preserve problem-relevant quantum numbers such as particle number, spin, or lattice momentum. The parameterized ansätze is constructed from the exponential map of the smaller Lie subalgebra $\mathfrak{su}(2^k)$ such that $ U(\boldsymbol{\theta}) = \exp\left( i \sum_{a=1}^{N^2-1} \theta_a \Lambda_a \right)$ with Pauli-monomial generators $ \Lambda_a \in \mathfrak{su}(2^k)$, whose dimensionality scales polynomially in system size. This formalism ensures that the reachable manifold is both expressive and low-dimensional, thereby eliminating the exponential suppression of gradient variance that characterizes barren plateaus. Furthermore, exploiting the generalized parameter-shift rule of Wierichs et al. \cite{wierichs2022general}, valid for arbitrary generators in $\mathfrak{su}(2^{k})$, we obtain exact analytic gradients using only additional forward-pass circuit evaluations, without finite-difference approximations or ancillary qubits.

The SUN-VQC framework provides a principled foundation for building variational circuits that are simultaneously trainable, symmetry-aware, and suitable for implementation on near-term quantum hardware. By confining the ansätze evolution to compact Lie subgroups tailored to the problem’s intrinsic symmetries, we retain algorithmic expressiveness while mitigating the curse of dimensionality. Numerical simulations on benchmark QML tasks—including molecular Hamiltonian learning, lattice spin model classification, and symmetry-conserving generative circuits—demonstrate that SUN-VQC maintains stable gradient magnitudes and significantly accelerates convergence relative to unconstrained baselines. Moreover, the method exhibits robustness against both coherent and incoherent noise sources, highlighting its compatibility with realistic device imperfections. Collectively, these results position SUN-VQC as a theoretically grounded and practically scalable approach for advancing quantum machine learning and quantum artificial intelligence.


\section{ Preliminary}
\subsection{\texorpdfstring{$\mathrm{SU}(N)$}{SU(N)} Gate Construction and Differentiable Optimization}

Variational-quantum-circuit behaviour is dictated by two closely related Lie–algebraic objects:  
(i) the \emph{dynamical Lie algebra} (DLA) generated by the control Hamiltonians actually implemented, and  
(ii) any \emph{embedded} Lie subalgebra that arises from an \textit{a priori} symmetry restriction such as acting only on a $k$-qubit block inside an $n$-qubit register.  
Making this distinction explicit eliminates the ambiguity noted in earlier drafts and provides a rigorous foundation for our barren-plateau analysis.

\textbf{Dynamical Lie algebra.}—For a finite-dimensional Lie group $G$ with Lie algebra $\mathfrak g=T_eG$, let  
$S=\{X_0,X_1,\dots,X_m\}\subset\mathfrak g$ denote the fixed set of control generators.  
The DLA is the smallest Lie subalgebra that contains~$S$,
\begin{equation}
\label{eq:DLA}
\mathfrak L_{\mathrm{dyn}}
        =\operatorname{Lie}(S)
        =\operatorname{Span}_{\mathbb R}\!
          \bigl\{\operatorname{ad}_{X_{i_1}}\!\cdots\!\operatorname{ad}_{X_{i_{k-1}}}(X_{i_k})\bigr\},
\end{equation}
where the adjoint actions are iterated over all finite words in the indices $i_j\in\{0,\dots,m\}$~\cite{wiersema2024classification,d2008lie}.

\textbf{Embedded symmetry blocks.}—Whenever the circuit acts only on a $k$-qubit subregister, an explicit embedding
\begin{equation}
\label{eq:embed}
\varphi:\;
SU\!\bigl(2^{k}\bigr)\longrightarrow SU\!\bigl(2^{n}\bigr),\qquad
U\mapsto\operatorname{diag}\!\bigl(U,\mathbbm 1_{\,2^{n-k}}\bigr),
\end{equation}
defines an embedded subgroup $H=\varphi\bigl(SU(2^{k})\bigr)$ of dimension $4^{k}-1$ with Lie algebra
$\operatorname{Lie}(H)=\{\operatorname{diag}(X,0)\mid X\in\mathfrak{su}(2^{k})\}\cong\mathfrak{su}(2^{k})$.  
If the control set $S$ equals a basis of this embedded algebra then $\mathfrak L_{\mathrm{dyn}}=\operatorname{Lie}(H)$; otherwise the DLA can grow larger, up to the full $\mathfrak{su}(2^{n})$ in generic hardware-efficient circuits.

\textbf{Gradient-variance scaling.}—For any cost function $C(\boldsymbol\theta)$ evaluated on an ansatz state $|\psi(\boldsymbol\theta)\rangle=U(\boldsymbol\theta)|\psi_0\rangle$ whose DLA is $\mathfrak g$, the variance of single-parameter gradients satisfies the universal bound~\cite{cerezo2021cost}
\[
\operatorname{Var}\bigl[\partial_{\theta}C\bigr]=\mathcal O\!\bigl(\dim\mathfrak g\bigr)^{-1}.
\]
Hardware-efficient Pauli$+$CNOT layers yield $\dim\mathfrak g=4^{n}-1$ and thus the notorious $\mathcal O(4^{-n})$ barren-plateau decay~\cite{mcclean2018barren}.  
By contrast, restricting every block to $k$ qubits gives
$\mathfrak g=\bigoplus_{b=1}^{n/k}\mathfrak{su}(2^{k})$ with $\dim\mathfrak g=\mathcal O(n)$ and restores a merely polynomial $\mathcal O(n^{-1})$ suppression, in full agreement with the numerical results reported later.

\textbf{Canonical $\mathrm{SU}(N)$ blocks.}—All subsequent benchmarks in this paper
employ two-qubit blocks ($k=2$, hence $N=2^{k}=4$) drawn from three
equally expressive gate families.  While their geometric
parameterisations differ, the dynamical Lie algebra of every block
coincides with the embedded $\mathfrak{su}(4)$:

\begin{enumerate}[label=(\roman*),leftmargin=*,nosep]
\item \textit{SUN-VQC block:}\;
      $U_{\mathrm{SUN}}
        =\exp\!\bigl(i\sum_{a=1}^{15}\theta_a\Lambda_a\bigr)$,
      corresponding to a single geodesic step endowed with an isotropic
      quantum-Fisher metric. The mathematical framework underlying this construction is discussed in Ref.~\cite{petz1996monotone}.
\item \textit{Cartan‐prior block:}\;
      $U_{\mathrm{Cartan}}
        =K_1\exp\!\bigl(i\boldsymbol\alpha\!\cdot\!\sigma_z\!\otimes\!\sigma_z\bigr)K_2$,
      a $KAK$ factorisation that cleanly separates local and entangling
      directions within the Weyl chamber\cite{zhang2003geometric}.
\item \textit{Pauli-rotation block:}\;
      $U_{\mathrm{Pauli}}
        =\prod_{m=1}^{15}\exp\!\bigl(i\theta_m P_m\bigr)$,
      an ordered product that still spans $\exp\!\bigl(\mathfrak{su}(4)\bigr)$
      but induces an anisotropic parameter metric\cite{mcclean2018barren}.
\end{enumerate}

Each construction preserves the $\mathrm{SU}(N)$ symmetry exactly,
mitigates barren-plateau scaling by design, and admits an analytic
parameter-shift rule for unbiased gradient evaluation, thereby preparing
the ground for the empirical studies in
Secs.~\ref{sec:methodology}–\ref{sec:experiments}.

\subsection{Gate Layer Architecture}

Each SUN-VQC layer is defined as a unitary transformation generated by a weighted sum of Pauli-monomial operators drawn from a subalgebra $ \mathfrak{su}(N) $, where $ N = 2^k $ and $ k $ denotes the number of qubits within the local block. Specifically, the layer takes the form
\[
U(\boldsymbol{\theta}) = \exp\left( i \sum_{a=1}^{N^2 - 1} \theta_a \Lambda_a \right),
\]
where $ \{ \Lambda_a \} \subset \mathfrak{su}(N) $ are traceless Hermitian generators formed from $ k $-qubit Pauli strings, typically chosen to reflect the global symmetries of the target Hamiltonian or dataset. While physical execution on gate-based quantum hardware ultimately relies on decomposition into native gates (e.g., CNOTs and single-qubit rotations), all optimization is performed in the canonical $ \mathfrak{su}(N) $ basis, yielding a uniform, isotropic parameter space. This geometric uniformity improves optimization stability by avoiding parameter-space distortions that arise from sequential gate products. Importantly, the global geodesic structure of the unitary manifold is preserved, ensuring minimal-length evolution between quantum states with respect to the Fubini–Study metric\cite{schirmer2002constructive}.

\begin{figure}[!t]
  \centering
  \includegraphics[width=\linewidth]{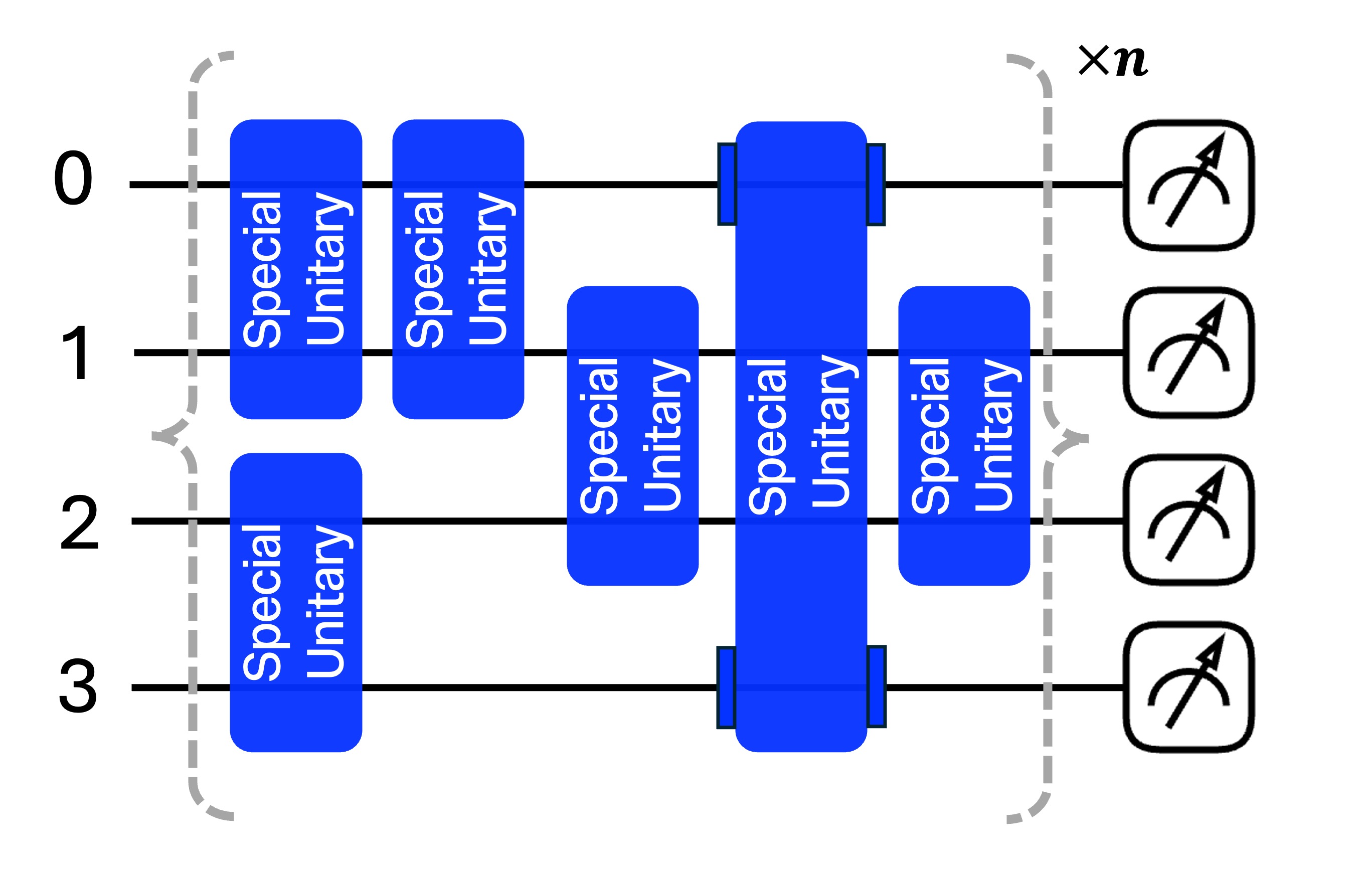}
  \caption{Brick-wall SUN-VQC on four qubits. Blue blocks are parametrised $\mathrm{SU}(4)$ two-qubit layers; alternating placement guarantees full connectivity. Qubits are read out in the computational basis.}
  \label{fig:sun-circuit}
\end{figure}

\subsection{Gradient Evaluation via Generalized Parameter-Shift Rule}

Efficient optimisation of SUN-VQC hinges on the ability to obtain exact gradients of expectation values with respect to the Pauli coordinates $ \boldsymbol{\theta} $ that parameterise each special unitary layer.  Consider a single coordinate $ \theta_a $ that couples to a Hermitian generator $ \Lambda_a \in \mathfrak{su}(N) $ through the exponential map $U_a(\theta_a)=\exp(i\theta_a\Lambda_a)$.  Writing the variational state as $ |\psi(\boldsymbol{\theta})\rangle = U(\boldsymbol{\theta})|\psi_0\rangle $ and an arbitrary observable as $ \hat{O} $, one may express the partial derivative of the expectation value in the Heisenberg picture,  
\[
\frac{\partial}{\partial\theta_a}\langle\hat{O}\rangle
      =\langle\psi_0|\,\nabla_{\theta_a}\!\bigl(U^\dagger(\boldsymbol{\theta})\hat{O}U(\boldsymbol{\theta})\bigr)\,|\psi_0\rangle .
\]
Because $ \Lambda_a $ is diagonalizable with a finite spectrum $ \{\lambda^{(a)}_j\}_{j=1}^N $, the Baker–Campbell–Hausdorff expansion shows that the above derivative can always be written as a finite linear combination of \emph{shifted} expectation values evaluated at parameter offsets that depend only on spectral gaps\cite{schuld2019evaluating}.  Let $ \Delta_{jk}^{(a)}=\lambda^{(a)}_j-\lambda^{(a)}_k $ denote the set of non-zero gaps; then a generalised shift formula,
\[
\frac{\partial}{\partial\theta_a}\langle\hat{O}\rangle
      =\sum_{r=1}^{R_a} c_{a,r}\Bigl[\,
      \langle\hat{O}\rangle_{\theta_a+s_{a,r}}
      -\langle\hat{O}\rangle_{\theta_a-s_{a,r}}\Bigr],
\]
holds with at most $R_a\leq|\{\Delta_{jk}^{(a)}\}|$ terms, where the coefficients $c_{a,r}$ and shifts $s_{a,r}$ are determined algebraically from the eigenvalue differences.  For generators possessing only two distinct eigenvalues, such as Pauli operators with spectrum $\{\pm1\}$, the rule collapses to the familiar two-term recipe with shifts $s=\pm\pi/2$; for higher-rank elements of $\mathfrak{su}(N)$ the number of terms grows at most linearly with the rank of the associated root system, guaranteeing a polynomial sampling overhead.

A concrete realisation of an $\mathfrak{su}(N)$ variational block is obtained by exponentiating a linear combination of Pauli words,
\[
U(\boldsymbol{\theta})=\exp\!\Bigl(i\sum_{m=1}^{d}\theta_m P_m\Bigr),\qquad
d=4^{k}-1,\;N=2^{k},
\]
where $\{P_m\}$ denotes the lexicographically ordered Pauli basis acting on a local register of $k$ qubits.  
Because each coefficient multiplies a Hermitian generator with a discrete spectrum, the derivative of any expectation value $f(U)$ with respect to $\theta_\ell$ can be written in closed form as 
\[
\frac{\partial}{\partial\theta_\ell}f\bigl(U(\boldsymbol{\theta})\bigr)
     =\sum_{m=1}^{d}\,i\omega_{\ell m}\Bigl[
        f\!\bigl(e^{-i\frac{\pi}{4}P_m}U\bigr)
       -f\!\bigl(e^{\,i\frac{\pi}{4}P_m}U\bigr)\Bigr],
\]
where the weights $\omega_{\ell m}$ are determined purely by the Lie-bracket relations of the Pauli basis and are independent of the circuit parameters\cite{wierichs2022general}.  
In the SUN-VQC setting the index range $d$ scales only polynomially with the global qubit count because the Pauli list is restricted to the chosen $\mathfrak{su}(N)$ irrep, not the full $\mathfrak{su}(2^{n})$.  
Consequently, every gradient component is obtainable with a constant—or at worst modestly growing—number of circuit evaluations, free from finite-difference bias and naturally amenable to batched execution on present-day quantum processors.  
This analytic, symmetry-aware differentiation rule underpins the scalable, geometry-consistent optimisation routine that lies at the heart of SUN-VQC.

\section{Variational Architecture and Expressivity Benchmark}\label{sec:methodology}

\subsection{Quantum Autoencoding Benchmark}
We address the quantum auto–encoding map 
$\mathcal H_{2}^{\otimes 6}\!\to\!\mathcal H_{2}^{\otimes 3}$, realised by a
variational unitary $U(\boldsymbol\theta)\in\mathrm{SU}(2^{6})$ acting on an
arbitrary six-qubit input $\ket{\psi}$.  
Defining
$\rho_{\text{trash}}(\boldsymbol\theta)=
  \operatorname{Tr}_{1,2,3}\!\bigl[U(\boldsymbol\theta)\ket{\psi}\bra{\psi}
  U^{\dagger}(\boldsymbol\theta)\bigr]$,
the figure of merit is the fidelity with the computational vacuum,
\begin{equation}
F(\boldsymbol\theta)=\bra{000}\rho_{\text{trash}}(\boldsymbol\theta)\ket{000},
\end{equation}
and its complementary infidelity
$L(\boldsymbol\theta)=1-F(\boldsymbol\theta)$.
For any fixed $\ket\psi$ the Schmidt decomposition across the
$3|3$ partition gives a rigorous lower bound
$L\ge L_{\mathrm{opt}}$, where
\begin{equation}\label{eq:Lopt}
L_{\mathrm{opt}}=1-\lambda_{\max},\qquad
\lambda_{\max}=\max_{j}\sigma_{j}^{2},
\end{equation}
with $\{\sigma_{j}\}$ the singular values of the $8\times8$ reshaping of
$\ket\psi$.

\subsection{Circuit Ansätze and Lie-Theoretic Structure}\label{ssec:ansatz_detail}
Each candidate circuit is built from an elementary \emph{two-qubit
block} $W(\boldsymbol\vartheta)\subset\mathrm{SU}(4)$ that is applied in a
depth-two brick-wall arrangement covering all nearest-neighbour pairs;
this yields a six-qubit unitary with $12$ blocks and
$180$ real parameters.  
Below we characterise the blocks from the standpoint of Lie theory and
quantum control.

\paragraph*{(i) SUN-VQC block.}
The most expressive architecture takes a single exponential of the full
Lie algebra $\mathfrak{su}(4)$,
\begin{equation}
W_{\text{SUN}}(\boldsymbol\vartheta)=
      \exp\!\Bigl(i\sum_{a=1}^{15}\vartheta_{a}\Lambda_{a}\Bigr),
\end{equation}
where $\{\Lambda_{a}\}$ is any orthonormal basis of
$\mathfrak{su}(4)$, e.g.\ the normalised Gell-Mann matrices.
The exponential map is surjective for $\mathrm{SU}(4)$, so every
two-qubit unitary is reachable in a \emph{single} step; moreover,
$\boldsymbol\vartheta$ provides a system of normal coordinates centred at
the identity, making the Fisher information isotropic to linear order.
Because the block acts on disjoint qubit pairs, the dynamical Lie algebra
of the entire six-qubit circuit is the direct sum
${\mathfrak g}^{(12)}\!=\bigoplus_{j=1}^{12}\mathfrak{su}(4)$,
whose dimension grows only linearly with problem size
and thereby suppresses barren-plateau scaling\cite{wiersema2024here,ragone2024lie}.

\paragraph*{(ii) Cartan-decomposition block.}
Any $V\in\mathrm{SU}(4)$ admits a
$KAK$ factorisation $V=K_{1}A K_{2}$ with
$K_{1,2}\!\in\!\mathrm{SU}(2)\otimes\mathrm{SU}(2)$
and
$A=\exp\!\bigl(i\,\sum_{\mu=1}^{3}\alpha_{\mu}\,\sigma^{\mu}_{z}\otimes
              \sigma^{\mu}_{z}\bigr)$
lying in the Cartan subalgebra\cite{yen2021cartan}.
We promote the six local angles in $K_{1,2}$ together with the three
Cartan angles $\boldsymbol\alpha$ to independent parameters,
retaining the full $15$-dimensional tangent space yet separating local
and non-local resources\cite{khaneja2001time}.
Geometrically, tuning $\boldsymbol\alpha$ moves the circuit inside the
Weyl chamber of two-qubit entanglers, while the $K$ factors explore the
isotropy subgroup that leaves the chamber invariant.
The block therefore spans the same Lie algebra
$\mathfrak{su}(4)$ as SUN-VQC but with an explicit geometric prior that
penalises unnecessary traversal of local directions.

\paragraph*{(iii) Pauli-rotation block.}
Here we order the $15$ non-identity two-qubit Pauli words
$\{P_{m}\}$ and apply them sequentially,
\[
W_{\text{Pauli}}(\boldsymbol\vartheta)=\prod_{m=1}^{15}
          \exp\!\bigl(i\vartheta_{m}P_{m}\bigr).
\]
Because the BCH expansion\cite{barenco1995elementary} closes on the Pauli set,
$\log W_{\text{Pauli}}\in\mathfrak{su}(4)$ and the reachable set equals
$\exp\bigl(\mathfrak{su}(4)\bigr)$.
However, the ordered product induces a highly anisotropic Riemannian
metric on parameter space: directions updated later in the product act on
a state that is already strongly entangled by previous rotations,
resulting in a broad distribution of gradient variances and a bias
towards shallow effective depths.

\paragraph*{(iv) Hardware-efficient block.}
The baseline employs the familiar pattern
\begin{equation}
\begin{aligned}
W_{\text{HE}}(\boldsymbol\vartheta) =\;&
\text{CNOT}_{1\to2} \\
&\times\, \bigl[R_{z}(\vartheta_{4}) R_{y}(\vartheta_{5}) R_{z}(\vartheta_{6})\bigr]_{2} \\
&\times\, \bigl[R_{z}(\vartheta_{1}) R_{y}(\vartheta_{2}) R_{z}(\vartheta_{3})\bigr]_{1} 
\end{aligned}
\end{equation}
which contains six local angles followed by a CNOT entangler.
While any sequence of such layers generates the full
$\mathfrak{su}(2^{n})$ under adjoint action, the accessible manifold
after a \emph{fixed} finite depth is a thin subvariety whose dimension
fails to grow with the number of layers once local rotations saturate.
Consequently, gradient norms decay exponentially with system size, a
signature of barren plateaus predicted by Lie-algebraic variance bounds.

\paragraph*{Comparative outlook.}
The first three ansätze implement single-step
$\exp(\mathfrak g)$ evolutions with \emph{stationary} Lie algebras
$\mathfrak g\subseteq\mathfrak{su}(4)$; optimisation therefore proceeds
on a homogeneous Riemannian manifold whose volume is polynomial in the
number of qubit pairs.
By contrast the hardware-efficient circuit explores a non-stationary
family of submanifolds obtained by alternating conjugation with CNOT
gates, leading to exponentially large search spaces and amplitude
concentration.  The numerical results in
Fig.~\ref{fig:learningcurves} corroborate this Lie-theoretic expectation:
SUN-VQC converges fastest, the Cartan prior imposes a mild slowdown yet
maintains favourable scaling, the Pauli-ordered product incurs a further
penalty due to metric anisotropy, and the hardware-efficient strategy
stagnates at the highest residual error.

\subsection{Gradient Evaluation and Optimisation}
Expectations $\partial_{\theta_{a}}F$ are evaluated with the
generalised parameter-shift rule for $\mathrm{SU}(N)$ generators.
For rank-one Pauli generators the rule collapses to the familiar
two-point form with $\pm\pi/2$ shifts, while Cartan generators of higher
rank demand at most a four-point stencil.
A fixed-step gradient descent with learning rate
$\eta=2\times10^{-2}$ is applied for $T=30,000$ iterations, a regime that
guarantees monotone convergence for every ansatz without invoking adaptive
schedules.

\subsection{Ensemble Averaging and Statistical Robustness}
\begin{figure}[!b]
    \centering
    \includegraphics[width=\linewidth]{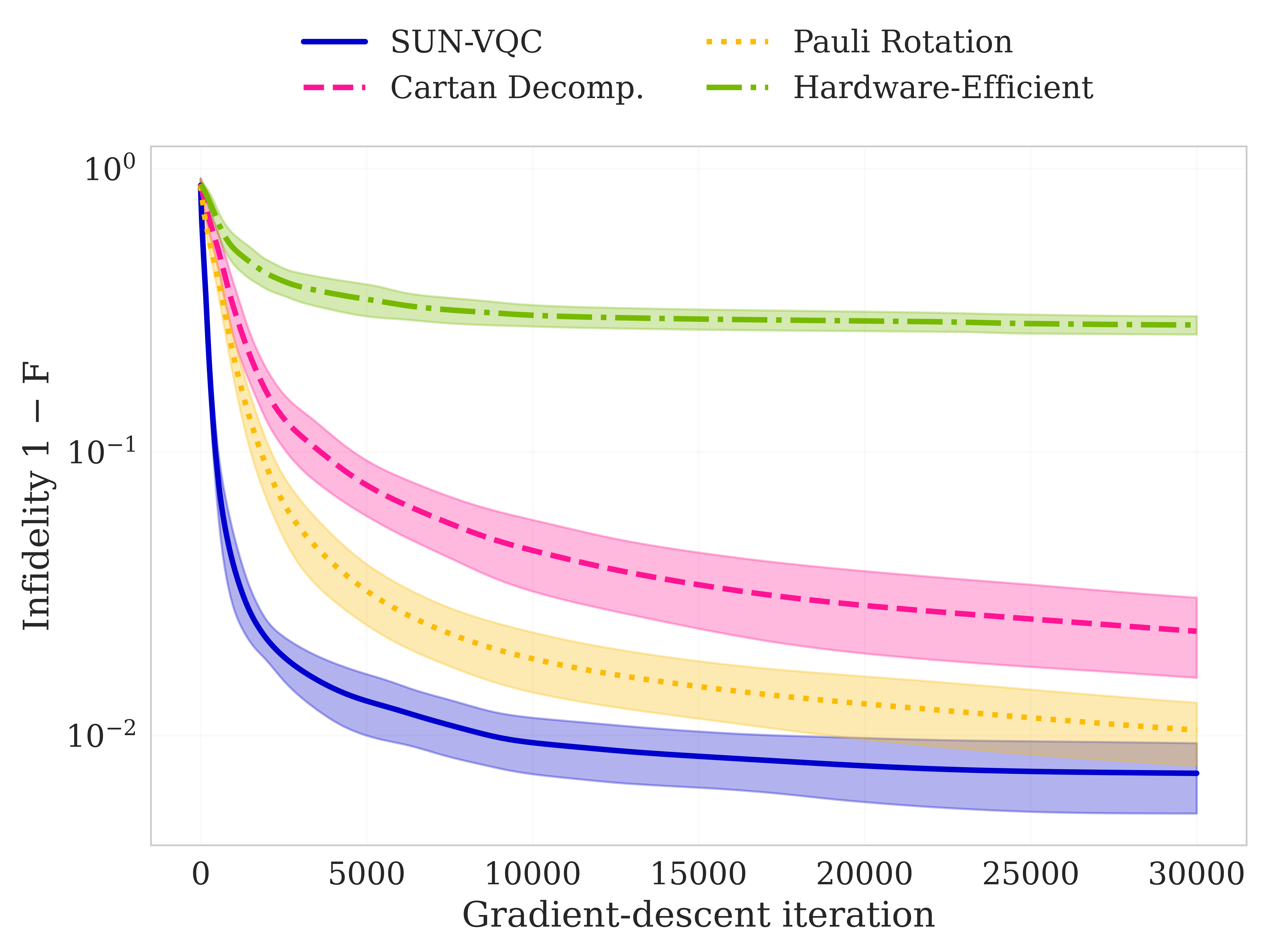}
    \caption{\label{fig:learningcurves}
    Mean infidelity $\bar L(t)$ for the four ansätze described in
    Sec.~\ref{sec:methodology}, averaged over
    $20$ Haar-random input states.
    Shaded regions indicate $\pm1\sigma$ across the ensemble.
    The SUN-VQC ansatz (solid blue) attains the lowest steady-state loss
    and exhibits the steepest initial descent,
    reflecting the favourable trade-off between expressivity and
    barren-plateau mitigation achieved by locality-preserving
    $\mathfrak{su}(4)$ blocks.}
\end{figure}

Robustness is assessed over an ensemble of
$N=20$ Haar-random inputs $\{\ket{\psi_{r}}\}_{r=1}^{20}$.
For each ansatz the sample mean
$\bar L(t)=\frac1N\sum_{r}L_{r}(t)$
and one standard deviation
$\sigma(t)$ are reported, with error bands plotted as
$\bar L\pm\sigma$ at $30\,\%$ opacity.  
The ensemble average of the theoretical bound~\eqref{eq:Lopt}
appears as a horizontal guideline in Fig.~\ref{fig:learningcurves}.

Fig.~\ref{fig:learningcurves} displays the typical learning curves on a
logarithmic infidelity scale.  
The SUN-VQC ansatz reaches the $10^{-2}$ fidelity threshold
within $\mathcal O(10^{3})$ iterations and continues to decrease
systematically, closely approaching the optimal bound.
This fast convergence is attributed to two complementary Lie-algebraic
features: (i) exact exploration of the full local algebra
$\mathfrak{su}(4)$ ensures expressivity, while
(ii) the restriction to pairwise blocks limits the
dynamical Lie algebra of the \emph{entire} circuit to a direct sum of
at most $12$ copies of $\mathfrak{su}(4)$,
thereby avoiding the exponential-dimensional search space
$\mathfrak{su}(64)$ encountered by generic hardware-efficient layers.
The Cartan decomposition, which fixes the non-local content to a single
Cartan element, converges more slowly yet maintains a clear advantage
over the Pauli-rotation sweep, underscoring the importance of
geometrically balanced parameterisation.
The purely hardware-efficient circuit stagnates
two orders of magnitude above the optimum; this plateau is consistent
with theoretical predictions that
layer-wise CNOT constructions suffer from
gradient suppression once their adjoint-generated Lie algebra
has saturated $\mathfrak{su}(2^{6})$.

\section{SUN-VQC-based Variational Quantum Classifier}
\label{sec:experiments}

\begin{figure}[!t]
    \centering
    \includegraphics[width=\columnwidth]{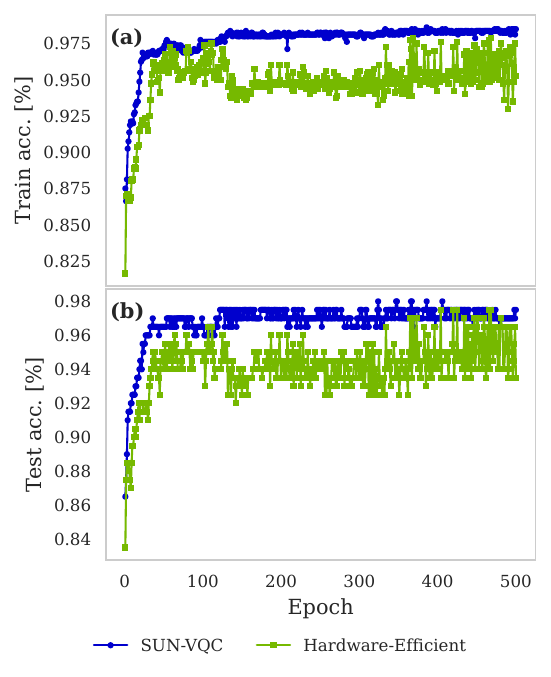}
    \caption{\label{fig:vqc_benchmark}
    Learning dynamics of the SUN-VQC (blue) versus
    an equally deep hardware-efficient ansatz (green) on the
    two-moons classification task.  Panel (a) shows the training accuracy;
    panel (b) shows test accuracy.  SUN-VQC attains approximately 98\% both training and test accuracy after 500 epochs, surpassing the Hardware-Efficient baseline by about 3 percentage points while demonstrating consistently lower variance throughout training.
    }
\end{figure}

In this section, we study a potential application using the case of a variational quantum classifier\cite{yano2021efficient}. The performance contrast in Fig.~\ref{fig:vqc_benchmark}
concisely captures the practical advantage of special-unitary
parameterisation. For this benchmark we encode the two-dimensional
\textit{two-moons} dataset into a four-qubit register and compare a
two-layer SUN-VQC built from brick-wall $\mathrm{SU}(4)$ blocks against an
equally deep hardware-efficient ansatz composed of $R_y$ rotations
followed by a nearest-neighbour CNOT ring.
Within the first $\sim50$ epochs the SUN-VQC already crosses the 95\%
accuracy threshold, whereas the hardware-efficient circuit approaches
that level only after $\sim200$ epochs and ultimately plateaus at
$\simeq96$.
The smaller dynamical Lie algebra
$\mathfrak g=\mathfrak{su}(4)^{\oplus2}$ explored by the SUN blocks
keeps gradient norms polynomial in system size, preserving optimisation
signal and enabling a \emph{monotonic} reduction of generalisation error.
Because both models employ identical classical post-processing and total
two-qubit depth, the gain is attributable solely to the
SU(4)-equivariant quantum layers rather than classical expressivity or
circuit depth.

Taken together with the auto-encoding benchmarks of
Sec.~\ref{sec:methodology}, these results confirm that symmetry-aware
Lie-subalgebra restriction is an effective barren-plateau mitigation
strategy that translates into tangible accuracy gains on noisy,
non-trivial data.
Crucially, the SUN-VQC achieves this without sacrificing hardware
feasibility: a single parametric SU(4) block can be realised as one
calibrated two-qubit pulse on today’s tunable-coupler or cross-resonance
platforms, suggesting an immediate path to device-level validation.

\section{Conclusion and Future Work}
\label{sec:discussion}
In summary, our results demonstrate that imposing a principled $\mathrm{SU}(N)$ equivariance on variational layers furnishes a concrete, symmetry-preserving path around the barren-plateau barrier: SUN-VQCs retain polynomial-scale dynamical Lie algebras, sustain gradient variance, and deliver stable accuracy on both unsupervised (auto-encoding) and supervised (two-moons) benchmarks without increasing physical depth. A natural caveat is that each exponential $\exp\left(i \sum_a \theta_a \Lambda_a\right)$ acts on $k$-qubit subblocks; when transpiled to native one- and two-qubit gates, this can still translate into non-trivial compilation overhead and hardware noise that may offset part of the theoretical advantage on today’s devices.
Looking forward, three avenues merit immediate exploration. (i) Scaling studies—analytically bounding, then empirically validating, gradient variance and sample complexity as we embed SUN blocks in lattices of $>50$ qubits will clarify the crossover where noise, rather than optimisation, becomes rate-limiting. (ii) Hardware implementation—deploying a single-pulse, parametric $\mathrm{SU}(4)$ gate on superconducting tunable-coupler or trapped-ion platforms will test whether the predicted variance advantage survives calibration errors and crosstalk. (iii) Generalised symmetry stacks—extending our framework to non-compact or projective groups (e.g., $\mathrm{Sp}(2N)$ for fermionic parity, or coset-space ansätze tailored to gauge constraints) could unlock barren-plateau-resilient circuits for quantum chemistry and lattice-gauge simulations.  By uniting Lie-algebraic control theory with variational learning, SUN-VQCs thus chart a scalable roadmap from proof-of-concepts to symmetry-aware quantum processors that learn efficiently at system sizes relevant to real-world quantum advantage.

\section*{Acknowledgment}
This work was supported by the Engineering and Physical Sciences Research Council (EPSRC) under grant number EP/W032643/1 and Imperial Quantum ICoNYCh Seed Fund.

\bibliographystyle{ieeetr}
\bibliography{references}

\end{document}